\documentclass[12pt]{iopart}
\usepackage{graphicx}
\usepackage{epstopdf}
\usepackage{amsfonts}
    
\begin{document}

   \title{The `twin paradox' in relativistic rigid motion}

   \author{Uri Ben-Ya'acov}

   \address{School of Engineering, Kinneret Academic College on
   the Sea of Galilee, \\   D.N. Emek Ha'Yarden 15132, Israel}

   \ead{uriby@kinneret.ac.il}

\vskip 2.0cm

\begin{abstract}
Relativistic rigid motion suggests a new version for the so-called `twin paradox', comparing the ages of two astronauts on a very long spaceship. Although there is always an instantaneous inertial frame in which the whole spaceship, being rigid, is simultaneously at rest, the twins' ages, measured as the proper-times along their individual world lines, are different when they are located at remote parts of the spaceship. The age, or proper-time, difference depends on the distance at rest between the astronauts and the rapidity difference between start to end. The relation of the age difference with the relative Doppler shift of light signals transmitted between the astronauts, and implications for the possibility to assign common age (proper-time) to complex, spatially extended, relativistic systems, are also discussed. The condition for simultaneous arrival of light signals emitted simultaneously from the opposite ends of a rigidly accelerating spaceship is resolved. \footnote{This article is a combination of recently published  paper and its addendum \cite{EJP2016a,EJP2016b}}
\end{abstract}

\pacs{03.30.+p}

\noindent{\it Keywords\/} : {twin paradox, relativistic rigid motion, proper-time, relativistic age, extended relativistic systems, rapidity}

\vskip30pt

\section{Introduction}

The notorious `twin paradox' served, from the early days of relativity theory, to illustrate and elucidate what seemed to be the bizarreness of the theory in contrast with daily experience. As is well known, the classical `twin paradox' (or `clock paradox' as a more formal title) uses a round-trip scenario, comparing the proper-time lapses as measured along two different world-lines between the same events in which these world-lines intersect (the spaceship's takeoff and eventual return to Earth). In this way the association between clock reading, age -- biological and physical age -- and proper-time lapses was established, both conceptually and empirically ({\it e.g.}, Rindler \cite{Rindler2006} p.64).

\begin{figure}[h]
  \centering
  \includegraphics[width=5cm]{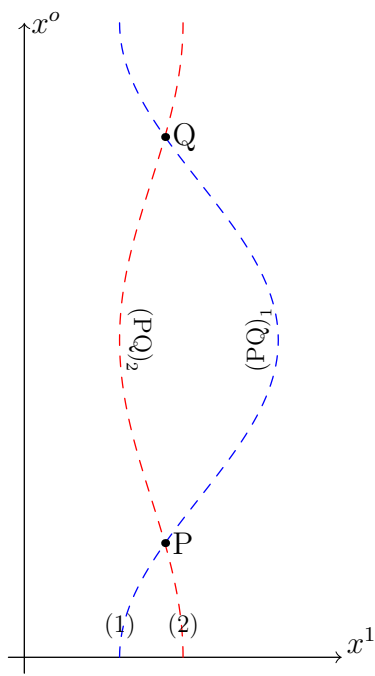}
  \begin{minipage}[b]{250pt}
  \caption{\label{fig: intersections} Proper-times may be compared only between intersections of world-lines.}
  \end{minipage}
\end{figure}

\begin{figure}[h]
  \centering
  \includegraphics[width=5cm]{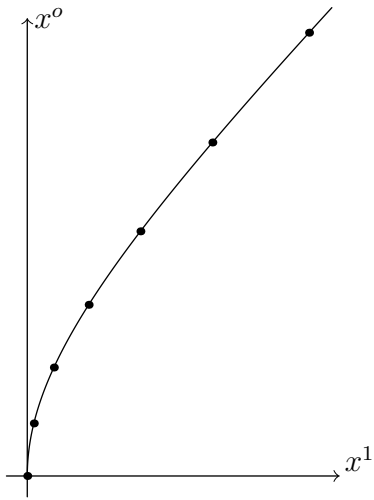}
  \begin{minipage}[b]{250pt}
  \caption{\label{fig: single} Space-time diagram showing an accelerated space-traveler's world-line as viewed relative to the Earth's rest frame, starting from rest. The intervals between neighbouring dots correspond to equal proper-time lapses, but due to time dilation they seem to be growing larger.}
  \end{minipage}
\end{figure}

Proper times may be defined and computed, therefore measured, as the Lorentz-invariant Minkowskian length of intervals on time-like world-lines. Comparison of the proper-time lapses measured along two different world-lines can only be done between common events, intersections of these world-lines. Let ${\rm P}$ and ${\rm Q}$ be such intersections, and $\left({\rm PQ}\right)_i \, , \, i=1,2$ the world-line intervals between these events along the different world-lines (see \Fref{fig: intersections}). Then the  proper-time lapses to be compared are the corresponding lengths of $\left({\rm PQ}\right)_{1,2}$ \footnote{It should be pointed out that trying to estimate proper-time relations just by looking at the diagrams may be misleading, because we are used to see Euclidean relations, while Minkowski space-time is pseudo-Euclidean. \Fref{fig: single}, for instance, demonstrates equal proper-time intervals that look to us larger and larger with growing velocity. On the other hand, world-line intervals that seem to be of equal lengths may correspond to different proper-time lapses, the higher the speed the shorter the proper-time lapse.}.  Both intervals cannot be geodesic, since that would mean that the intervals coincide. Therefore at least one of the world-lines must be nongeodesic. In the classical round-trip scenario one world-line is geodesic (inertial) while the other is not -- it is accelerated, and being nongeodesic the corresponding proper-time lapse on it is shorter. But other scenarios, involving two nongeodesic world-lines, are possible.


Time-like world-lines correspond to point-like particles. Point-like particles are idealizations -- in reality we have extended systems whose different points move on different world-lines to which would correspond, in general, different proper-times, resulting in differential ageing within the system. How, then, can we discuss differential ageing in such systems ? Is it possible to assign a common or representative proper-time for the whole system that may serve as its age ?

To illustrate the issue, let us consider the following alternative version for the `twin paradox' :

\begin{quote}
 \textit{In this story, both twins are astronauts, assigned for the same space mission, which is about to launch in a very long spaceship. One of them is located at the front end of the spaceship, the other at its rear end. While on Earth, the brothers surely are the same age. The spaceship starts from rest, and after a while in space lands on a remote planet somewhere in the galaxy. As the spaceship comes to rest the brothers walk towards the centre of the spaceship where they meet. Will they still be of the same age ?}
\end{quote}

This is a true `twin story', since after the journey along different world-lines the brothers meet and can compare their ages. Unlike the classical scenario, in which the traveling twin's spaceship may be assumed point-like, here the spaceship must be an extended body. It is therefore assumed that the spaceship is so designed that it remains rigid all through the voyage. The reason for this assumption is that the relativistic rigidity condition, first suggested by Born in 1909 \cite{Born1909}, requires all the parts of the body or system in question to be in any moment at rest relative to a common momentary inertial frame. Another way to put it is that the distance between any two particles in the system must remain unchanged throughout the motion relative to an instantaneous inertial rest frame attached to any one of the two particles -- a very reasonable requirement for a spaceship in voyage.

How does rigidity determine the answer ?

Immediately after its inception, it was asserted by Herglotz and Noether \cite{HerglotzNoether1910} that Born's condition implies that  accelerated rectilinear rigid motion is possible. For this situation to be maintained, different points along the spaceship must have different accelerations which are inter-related in a very specific way. It is assumed in the following that the issue of differential acceleration was solved by the engine design of the spaceship, so we need not be bothered by it.

Most of the discussions of relativistic linear rigid motion assumed constant proper accelerations (see, {\it e.g.} Rindler \cite{Rindler2006} p.71, Franklin \cite{Franklin2012}, and references therein) implying hyperbolic motion. This is a convenient but un-necessary limitation, because by the Herglotz-Noether theorem \cite{HerglotzNoether1910} the relativistic rigidity condition is satisfied, in rectilinear motion, also by time-varying accelerations. Moreover, time varying accelerations allow a variety of scenarios, from one in which the spaceship launches from rest in one inertial reference frame and lands, coming to rest, in another inertial reference frame, to a round-trip scenario in which the acceleration cannot remain constant throughout the whole journey.

General, with possibly time-varying accelerations, linear rigid motion was explicitly discussed, in part, by Kim and Jo \cite{KimJo2004}, but not in a Lorentz covariant manner, and not referring to proper-times and ageing, which is our main interest here. Age, like proper (rest-)mass, is an object's intrinsic property, and should therefore be treated in a Lorentz covariant manner. We therefore start (\Sref{sec: LRRM}) considering linear relativistic rigid motion with general (not-necessarily constant) accelerations in fully Lorentz covariant notation, which allows us to relate accelerations, velocities and proper-times of arbitrarily different points along the moving body. With these relations the proper-times of the two brothers are compared (\Sref{sec: twinast}). Differential ageing is computed, found to be proportional to the proper spatial distance between the two and to the rapidity difference between start to end. Therefore, if the end station is moving relative to the home station, then the brothers do indeed end up with different ages, simply because of being located in remote parts of the spaceship. The transmission of signals between the astronauts is examined (\Sref{sec: sigtrans}), establishing a relation between the relative Doppler shift of these signals and the age difference. The issue of simultaneous arrival of simultaneously emitted signals is also discussed. The paper is concluded (\Sref{sec: conrem}) with implications for the possibility of assigning a common proper-time to complex relativistic systems, and a comment on the relation between the rigidity condition and simultaneity.

The convention $c=1$ is used throughout, except for equation \eref{eq: twinsage}. Events in Minkowski space-time are $x^\mu = \left(x^o,x^1,x^2,x^3\right)$, and the metric tensor with positive signature $g_{\mu\nu} = {\rm diag} \left(-1,1,1,1\right) \, , \, \mu,\nu = 0,1,2,3$, is assumed. For any 4-vectors $a^\mu$ and $b^\mu$, their inner product is $a \cdot b = g_{\mu\nu} a^\mu b^\nu$ using Einstein's summation convention.

\vskip20pt

\section{Linear relativistic rigid motion with general accelerations} \label{sec: LRRM}

We start by discussing rectilinear rigid motion with time-varying proper accelerations in a Lorentz-covariant manner.

To analyze the rigid motion of the spaceship it is convenient to choose an arbitrary reference point within the system. It defines a reference world-line $x^\mu = x_o^\mu \left( \tau _o \right)$ with $\tau _o$ its proper-time, with the unit velocity 4-vector $u_o^\mu  = \dot x_o^\mu \left( \tau _o \right)$, the over-dot implying differentiation relative to $\tau _o$. We recall that a proper-time element along a general world-line $x^\mu \left(\tau\right)$ is defined via the Minkowskian line element $d\tau = \sqrt{ - dx \cdot dx}$. Alternatively, the proper-time $\tau$ may be defined as the time-like parameter for which the 4-velocity $u^\mu  = d x^\mu / d \tau$ is always a unit 4-vector ($u \cdot u = -1$).

It should be pointed out that although the following derivation requires a reference point, its choice is completely arbitrary, as is verified in the following, in particular in equations \eref{eq: aa-ao} - \eref{eq: aa-ab2}.

At each point $x_o^\mu \left( \tau _o \right)$ on the reference world-line an orthonormal spatial triad $n_i^\mu \left( \tau _o \right), \, i = 1,2,3,$ may be defined, spanning the 3-space orthogonal to $u_o^\mu \left( \tau _o \right)$, thus defining a 3-D reference frame attached to the spaceship. This is the simultaneity hyperplane relative to $x_o^\mu \left( \tau _o \right)$. Together the orthonormal tetrad $\left( u_o^\mu ,n_i^\mu  \right)$ is defined with the relations
 \begin{equation} \label{eq: ortho}
 n_i \cdot n_j = \delta _{ij}   \quad , \quad   u_o \cdot n_i = 0   \quad , \quad   u_o \cdot u_o =  - 1
 \end{equation}
Any other point in the system may be defined relative to the reference world-line by a set of 3 constant distance parameters $\left\{ \zeta_A^i \right\}$ relative to the triad $n_i^\mu \left( \tau_o \right)$, $A$ being an index designating the particular point, thus invoking the rigidity condition. The relative Minkowskian displacement $\xi_A^\mu\left( \tau_o \right)  = \zeta_A^i n_i^\mu$ which lies in the simultaneity hyperplane relative to $x_o^\mu \left( \tau_o \right)$ defines the world-line of the $A$-th point
 \begin{equation} \label{eq: xxi}
 x_A^\mu \left( \tau_o \right) = x^\mu \left( \zeta_A,\tau_o \right) = x_o^\mu \left( \tau_o \right) +\xi_A^\mu\left( \tau_o \right) = x_o^\mu \left( \tau_o \right) + n_i^\mu \left( \tau_o \right) \zeta_A^i
 \end{equation}

The motion of the whole system is completely determined by that of the reference world-line and the triad $n_i^\mu \left( \tau_o \right)$ attached to it. Without loss of generality, $n_1^\mu$ may be chosen in the direction of the 4-acceleration of the reference point, satisfying the relation $\dot u_o^\mu  = a_o^\mu  = a_o n_1^\mu$. The scalar coefficient $a_o = \dot u_o \cdot n_1$  is the proper acceleration, $a_o > 0$ or $< 0$ when the spaceship accelerates or decelerates, respectively. From the orthonormality conditions \eref{eq: ortho} it follows that $\dot u_o \cdot n_1 = - u_o \cdot \dot n_1$. The condition for rectilinear motion (no spatial rotation) $n_i  \cdot \dot n_j = 0$ allows $\dot n_1^\mu$ to be directed only along $u_o^\mu$, so that necessarily $\dot n_1^\mu = a_o u_o^\mu$. The other two tetrad vectors $n_i^\mu \; (i=2,3)$ correspond to displacements perpendicular to the spatial direction of motion, and without rotation are constant. Therefore, finally, the equations of the tetrad $\left( u_o^\mu ,n_i^\mu  \right)$ are
 \begin{equation} \label{eq: fbeq}
 \dot u_o^\mu  = a_o^\mu  = a_o n_1^\mu  \quad , \quad  \dot n_1^\mu = a_o u_o^\mu  \quad , \quad  \dot n_i^\mu = 0 \quad i=2,3
 \end{equation}

The prime result of applying the rigidity condition for rectilinear motion is that all the points in the simultaneity hyperplane do indeed move with the same velocity, or, in other words, at each moment there is an instantaneous rest frame common to all the points of the system : From equation \eref{eq: fbeq} it follows that $\dot \xi_A^\mu  = \zeta_A^i \dot n_i^\mu  = \zeta_A^1 a_o u_o^\mu$. Then, with $\tau_A$ the proper-time at the $A$-th point, the unit 4-velocity there is
 \begin{equation} \label{eq: utau}
 u_A^\mu \left( \tau_A \right) = \left( \frac{d\tau_A}{d\tau_o} \right)^{-1} \frac{d}{d\tau_o} x_A^\mu\left( \xi_A,\tau_o \right) = \left( \frac{d\tau_A}{d\tau_o} \right)^{- 1}  \left( 1 + \zeta_A^1 a_o \right) u_o^\mu
 \end{equation}
Since both $u_A^\mu$ and $u_o^\mu$ are unit 4-velocities (with ${u_A}^2 = {u_o}^2 = -1$), it follows that the coefficient of $u_o^\mu \left( \tau_o \right)$ in the RHS of equation \eref{eq: utau} must be unity. Therefore $u_A^\mu \left( \tau_A \right) = u_o^\mu \left( \tau_o \right) $, and as a bonus we receive the relation between the proper-times,
 \begin{equation} \label{eq: tato}
 \frac{d\tau_A}{d\tau_o} = 1 + \zeta_A^1 a_o
 \end{equation}

Since the 4-velocities are identical at all the points in the simultaneity hyperplane, but not the proper-times, the accelerations are point-dependent :
 \begin{equation} \label{eq: ao-aa}
 a_A^\mu \left( \xi_A,\tau_o \right) = \frac{du_A^\mu}{d\tau_A} = \left( \frac{d\tau_A}{d\tau_o} \right)^{-1} a_o^\mu \left( \tau_o \right) = \frac{a_o }{1 + \zeta_A^1 a_o} n_1^\mu
 \end{equation}
Therefore, the accelerations at all the points are parallel (as expected, necessarily, for rectilinear motion) and entirely determined by the acceleration of the reference point. Yet, it is important to show that the choice of the reference point is completely arbitrary : Writing $a_A^\mu = a_A n_1^\mu$ the reciprocal relation is easily obtained from \eref{eq: ao-aa},
 \begin{equation} \label{eq: aa-ao}
 a_o  =  \frac{a_A }{1 - \zeta_A^1 a_A}
 \end{equation}
Any two points in the same simultaneity hyperplane then satisfy the identity
 \begin{equation} \label{eq: aa-ab}
 \frac{a_A }{1 - \zeta_A^1 a_A}  =  \frac{a_B }{1 - \zeta_B^1 a_B}
 \end{equation}
While this relation depends separately on the position parameters relative to the reference world-line $\zeta_A^1$ and $\zeta_B^1$, it is possible, with some basic algebraic steps, to derive from equation \eref{eq: aa-ab} another relation which depends only on the relative position of the two points, independent on the initial reference point :
 \begin{equation} \label{eq: aa-ab2}
 a_A  =  \frac{a_B}{1 + \left( \zeta_A^1 - \zeta_B^1 \right) a_B}
 \end{equation}
Therefore, any point can be chosen as the reference point with the same result -- there is no preferred point in the system.

Back to the proper-times relation \eref{eq: tato}, we now use a basic relation between the proper acceleration, the proper-time and the rapidity $\eta\left( v \right) \equiv  \tanh^{-1} \left( v \right)$ (the additive quantity in the superposition of co-linear velocities \cite{RhodesSemon2004}) : Consider a point particle moving linearly on the world-line $x^\mu = \left( t,x(t),y,z \right)$, with fixed $y,z$, relative to some inertial frame. With unit 4-velocity $u^\mu = \gamma(v) \left( 1,v,0,0 \right)$, with $v(t) = dx/dt$ and $\gamma(v) = \left( 1-v^2 \right)^{-1/2} = dt / d\tau$, its 4-acceleration is
 \begin{equation}\label{eq: acc}
 a^\mu  = \frac{du^\mu }{d\tau} = \gamma^4 \left( v \right) \frac{dv}{dt} \left( v,1,0,0 \right)
 \end{equation}
Since $n_1^\mu = \gamma \left( v \right)\left( v,1,0,0 \right)$ is a space-like unit 4-vector, the proper acceleration $a$, satisfying $a^\mu = a n_1^\mu$, is $a = \gamma^2 \left( v \right) \left( dv/d\tau \right)$. From the rapidity definition follows $d\eta = \gamma^2 \left( v \right) dv$, so that the relation
 \begin{equation}\label{eq: tauaeta}
 a  = \frac{d\eta }{d\tau} \, ,
 \end{equation}
which holds for all rectilinear motion, is obtained \cite{Minguzzi2005}.

Substituting eq. \eref{tauaeta} in \eref{eq: tato} therefore yields
 \begin{equation} \label{eq: tatoeta}
 d\tau_A = d\tau_o + \zeta_A^1 d\eta
 \end{equation}
Since all the points in any simultaneity hyperplane move with the same velocity, $\eta$ has the same value for all the spaceship parts on simultaneity hyperplanes. In fact, $\eta$ may be used to characterize and even parametrize the simultaneity hyperplanes. Proper-times are defined up to an additive constant. Therefore, assuming a starting simultaneity hyperplane where the proper-times are the same for all the spaceship points, equation \eref{eq: tatoeta} is integrated for the explicit relation between the proper-times at the end of the journey,
 \begin{equation} \label{eq: ta-to-vo}
 \tau_A = \tau_o + \zeta_A^1 \left[\eta \left( v_{\rm end} \right) - \eta \left( v_{\rm start} \right) \right]
 \end{equation}
This result is obviously independent of the choice of the reference point $x_o^\mu$ and the particular details of the acceleration, since it also follows from eq.\eref{eq: tatoeta} that the proper times at any two points on the spaceship later satisfy the relation
 \begin{equation} \label{eq: tatbeta}
 \tau_A - \zeta_A^1 \eta =  \tau_B - \zeta_B^1 \eta = \tau_o
 \end{equation}
Recalling that rapidity differences are Lorentz invariant insures the Lorentz invariance of these results.

Finally, since the ratio or relative advancement of proper-times must be positive, so that $d\tau_A/d\tau_o > 0$ for all points $A$ and all possible choices of the reference point, then follows from equation \eref{eq: tato} the condition $1 + \zeta_A^1 a_o > 0$ or $\left|a_o\right| < \left|\zeta_A^1\right|^{-1}$ for all $\zeta_A^1$. In particular, if $L$ is the spaceship's length and the reference point chosen at its centre, then the condition reads $\left|a_o\right| < 2/L$. In practice, this upper bound is very high : Writing $c$ explicitly, then even for $L = 1{\rm km}$ we get $\left|a_o\right| < 2c^2/L = 1.8 \times 10^{14} {\rm m/s^2}$.

\vskip20pt

\section{Differential ageing of the twin astronauts} \label{sec: twinast}

We are now ready to launch into the space voyage with the twin astronauts. Let the length of the spaceship be $L$, and let it start from rest while parking along a pier of the same length. Of the two astronauts, let A be positioned at the rear of the spaceship and B positioned at the front. A's world-line may serve as the reference world-line, written in terms of the home-base coordinates as
 \begin{equation}\label{eq: xA}
 x_{\rm{A}}^\mu = \left( t,x(t),0,0 \right)
 \end{equation}
Its unit 4-velocity is then $u^\mu = \gamma(v) \left( 1,v,0,0 \right)$, and initial conditions are assumed at $t = 0$ : $x(0) = 0 \, , \, v(0) = 0$. Following the relations in and around equation \eref{eq: acc}, the instantaneous simultaneity hyperplanes are defined by the space-like unit vector $n_1^\mu = \gamma \left( v \right)\left( v,1,0,0 \right)$. Then, with $\zeta^1 = L$, B's world-line is
 \begin{equation}\label{eq: xB}
 x_{\rm{B}}^\mu = x_{\rm{A}}^\mu + L n_1^\mu = \left( t + L \gamma(v) v,x(t) + L \gamma(v),0,0 \right)
 \end{equation}

\begin{figure}[h]
  \centering
  \includegraphics[width=6cm]{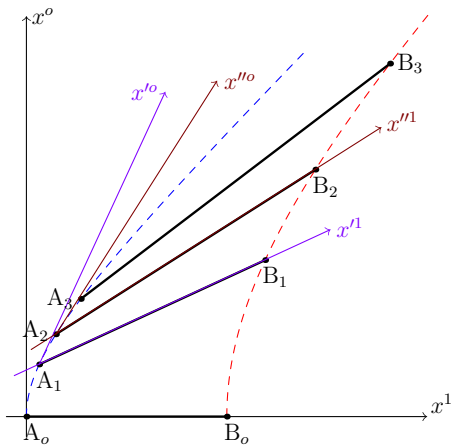}
  \begin{minipage}[b]{250pt}
  \caption{\label{fig: spaceship} Space-time diagram showing the spaceship's voyage as viewed from the Earth's rest frame $\left(x^o,x^1\right)$, starting from rest at  ${\rm A_o}{\rm B_o}$. The dashed lines show the world-lines of the astronauts. The bold lines ${\rm A_1}{\rm B_1}$,  ${\rm A_2}{\rm B_2}$, ${\rm A_3}{\rm B_3}$ show the position of the spaceship at some chosen moments during the voyage, with corresponding instantaneous rest frames $\left(x'^o,x'^1\right) \, , \, \left(x''^o,x''^1\right)$. Proper-times are measured as Minkowskian length of world-line intervals, {\it e.g.}, ${\rm A_o}{\rm A_1}$, ${\rm B_o}{\rm B_2}$, etc. The apparent spaceship's elongation to $\gamma(v) L$ is fictitious, due to Lorentz transformation from the spaceship's rest frame to the Earth's frame. The world-lines drawn using equation \eref{eq: xABhyp}.}
  \end{minipage}
\end{figure}

For each value of $t$, the events $x_{\rm{A}}^\mu(t)$ and $x_{\rm{B}}^\mu(t)$ correspond to different home-base times, but they are simultaneous relative to the spaceship (more precisely, they lie on the same instantaneous simultaneity hyperplane). Therefore, while $x_{\rm{A}}^o(t) = t$, for the simultaneous (relative to the spaceship) B-event $x_{\rm{B}}^o(t) = t + L \gamma[v(t)] v(t) \ne t$. Considering both world-lines together, then $t$ should be regarded merely as a time-like parameter. The identity of the velocities on the simultaneity hyperplane is verified by the relation
 \begin{equation}\label{eq: vB}
 v_{\rm{B}} = \frac{dx_{\rm{B}}^1}{dx_{\rm{B}}^o} = \frac{d\left[ x(t) + L \gamma(v) \right]}{d\left[ t + L \gamma(v) v \right]} = \frac{v + L v \gamma^3(v) (dv/dt)}{ 1 + L \gamma^3(v) (dv/dt) } = v
 \end{equation}
Both world-lines are shown in the space-time diagram in \Fref{fig: spaceship}.

As an illustration for these relations, then for hyperbolic motion the world-lines are (conveniently parametrized by the rapidity $\eta$)
\begin{equation}\label{eq: xABhyp}
\eqalign{
  x_{\rm{A}}^\mu &= \left( \frac{1}{a} \sinh\eta,\frac{1}{a} \left(\cosh\eta - 1\right),0,0 \right) \\
 x_{\rm{B}}^\mu &= \left( \frac{1+ a L}{a} \sinh\eta , \frac{1+ a L}{a} \cosh\eta - \frac{1}{a},0,0 \right)}
\end{equation}
$a$ is A's proper acceleration, while $a / \left(1 + aL\right)$ is B's proper acceleration, and the proper-times are $d\tau_{\rm A} = d\eta / a$ and $d\tau_{\rm B} = \left(1 + aL\right) d\eta / a$, respectively. Each value of $\eta$ defines an instantaneous rest frame (see \Fref{fig: spaceship}).

\begin{figure}[h]
  \centering
  \includegraphics[width=5cm]{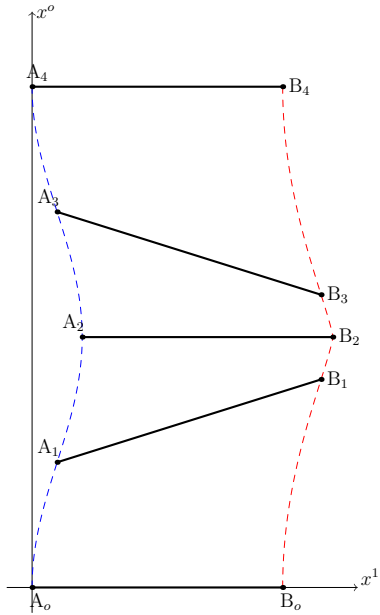}
  \begin{minipage}[b]{250pt}
  \caption{\label{fig: rtrip} Space-time diagram showing a spaceship's round trip as viewed from the Earth's rest frame $\left(x^o,x^1\right)$, starting from rest at ${\rm A_o}{\rm B_o}$ and returning to rest at ${\rm A_4}{\rm B_4}$. The change in the inclination of the spaceship's position is evident, due to reversing the direction of motion. Although the world-line intervals ${\rm A_o}{\rm A_4}$ and ${\rm B_o}{\rm B_4}$ look somewhat different, their Minkowskian lengths (proper-time lapses) are equal. This diagram uses $x(t) = 0.2 \sin^2(0.5 \pi t)$ with $L = 1$ to insure condition \eref{eq: ABcond}.}
  \end{minipage}
\end{figure}

According to \eref{eq: ao-aa} each point of the spaceship requires its own acceleration to maintain the assumed rigidity; as the spaceship accelerates (in the $+x$ direction) A suffers the highest acceleration, which gradually decreases along the spaceship in the direction of its motion. If the acceleration changes and the spaceship decelerates these relations reverse. Then we should be aware of the fact that due to the obvious condition $d\tau_{\rm A} / d\tau_{\rm B} > 0$ then follows from \eref{eq: tato} the condition
 \begin{equation} \label{eq: ABcond}
 1 + L \gamma^3(v) \frac{dv}{dt} > 0  \quad \Rightarrow \quad  \gamma^3(v) \frac{dv}{dt} > -\frac{1}{L}
 \end{equation}
in accordance with the discussion following equation \eref{eq: tatbeta}.

At the home-base the spaceship starts from rest, so that $v_{\rm{start}} = 0$ and both twins start at the same age, $\tau_{\rm{A}} \left({\rm start}\right) = \tau_{\rm{B}} \left({\rm start}\right)$. If the spaceship returns, eventually, to its home-base and lands there (as illustrated by \Fref{fig: rtrip}), then also $v_{\rm{end}} = 0$, and both twin's ages are equal. But if the spaceship arrives at a remote star system which moves with velocity $V$ relative to the home-base, then it follows from \eref{eq: tatbeta} that the twins' proper-times differ :
 \begin{equation} \label{eq: twinsage}
 \tau_{\rm{B}} \left( \rm{end} \right) = \tau_{\rm{A}} \left( \rm{end} \right) + \frac{L}{c} \tanh^{-1} \left( \frac{V}{c} \right)
 \end{equation}
(the light velocity $c$ is explicitly introduced in equation \eref{eq: twinsage} for the following computation). The proper-time difference is therefore determined in terms of the relative velocity between the two stations.

As the spaceship lands and comes to rest in the end station the astronauts' world-lines still do not intersect, them being situated at remote ends of the spaceship. But then they start walking towards each other (presumably with the same speed relative to the station's rest frame), so the proper-time lapses between landing and their meeting is the same for both, and doesn't change the proper-time difference \eref{eq: twinsage} which as we have now verified determines the age difference between the astronauts.

Although the difference is real, in practice it is very minute : Let the spaceship be 1km long and $V=0.9c$. Then
 \begin{equation} \label{eq: ptimediff}
 \Delta\tau \left( \rm{end} \right) = \tau_{\rm{B}} \left( \rm{end} \right) - \tau_{\rm{A}} \left( \rm{end} \right)  \approx 4.9 \times 10^{-6} {\rm sec}
 \end{equation}
The effect is real, but hardly detectable.

\vskip20pt

\section{Signal transmission during the journey} \label{sec: sigtrans}

An interesting by-product of the foregoing discussion relates transmission of EM signals between the astronauts and the differential ageing \eref{eq: twinsage}. Suppose that the twins, wishing to entertain themselves during the long journey, start exchanging signals. Let twin A, in the rear, send at some moment a signal to B in the front. We recall that at each moment there is an instantaneous inertial frame in which the spaceship is momentarily at rest. Let us denote the instantaneous rest frame that corresponds to emitting the signal ${\rm S}_{\rm o}$, and assume that the astronauts' world-line relative to ${\rm S}_{\rm o}$ are given by \eref{eq: xA} and \eref{eq: xB}, so $t$ is the time as measured by the ${\rm S}_{\rm o}$-clocks for A. Relative to ${\rm S}_{\rm o}$, then, the emission event may be assumed to be ${\rm A}_o = \left(0,0,0,0\right)$. Simultaneously relative to ${\rm S}_{\rm o}$, the other astronaut is at the event ${\rm B}_o = \left(0,L,0,0\right)$.

\begin{figure}[h]
  \centering
  \includegraphics[width=6cm]{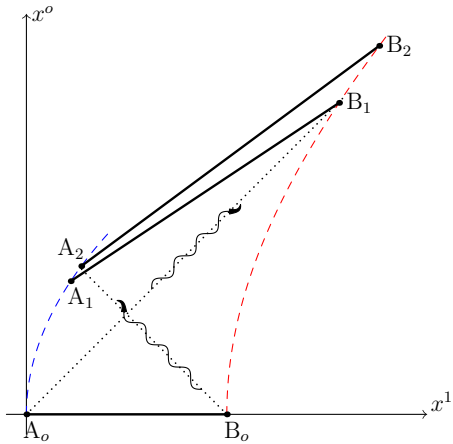}
  \begin{minipage}[b]{250pt}
  \caption{\label{fig: lightnohyp} Space-time diagram showing light signals sent from A (${\rm A}_o$) to B (${\rm B}_1$) and from B (${\rm B}_o$) to A (${\rm A}_2$). This diagram shows the general case, in which the signals, although emitted simultaneously, don't arrive simultaneously relative to the spaceship instantaneous rest-frame (${\rm A}_1{\rm B}_1 \ne {\rm A}_2{\rm B}_2$). The diagram uses $x(t) = 0.25t^2$, in accordance with example used in equation \eref{eq: simcondat2}.}
  \end{minipage}
\end{figure}

The signal moves in a straight line along the spaceship (assuming that the interior of the spaceship allows it a free path) and arrives at B, which is now moving with some velocity $v_1$ relative to ${\rm S}_{\rm o}$ due to the acceleration of the spaceship while the signal was traveling. The momentary rest frame now is different than ${\rm S}_{\rm o}$, and may be denoted ${\rm S}_1$. The event of the signal arrival to B may be denoted ${\rm B}_1$; according to \eref{eq: xB} it corresponds to some value $t = t_1$ of A's ${\rm S}_{\rm o}$-time, so that ${\rm B}_1  = \left( t_1 + L \gamma(v_1) v_1,x(t_1) + L \gamma(v_1),0,0 \right)$. Simultaneously relative to $S_1$, the other astronaut is at the event ${\rm A}_1 = \left(t_1,x(t_1),0,0\right)$. The light-cone condition for the signal implies
 \begin{equation} \label{eq: AtoB}
 t_1 + L \gamma(v_1) v_1 = x(t_1) + L \gamma(v_1)  \quad \Leftrightarrow \quad  t_1 - x(t_1) = e^{-\eta(v_1)} L
 \end{equation}
Since the signal was emitted when the spaceship was at rest relative to ${\rm S}_{\rm o}$, it arrives at B red-shifted, with the Doppler factor
 \begin{equation} \label{eq: DoppAB}
 \sqrt{\frac{1-v_1}{1+v_1}} = e^{-\eta(v_1)} = \exp{\left[\frac{\tau\left({\rm A}_o{\rm A}_1\right) - \tau\left({\rm B}_o{\rm B}_1\right)}{L}\right]}
 \end{equation}
where $\tau\left({\rm A}_o{\rm A}_1\right)$ and $\tau\left({\rm B}_o{\rm B}_1\right)$ are, respectively, the proper-time lapses along the corresponding world-lines intervals ${\rm A}_o{\rm A}_1$ and ${\rm B}_o{\rm B}_1$.

Similarly, if B sends at ${\rm B}_o$ a light signal to the back of the spaceship, the signal arrives to A at some $t = t_2$ when the spaceship is moving with velocity $v_2$ relative to ${\rm S}_{\rm o}$, so the arrival event is ${\rm A}_2 = \left(t_2,x(t_2),0,0\right)$ with the light-cone condition
 \begin{equation} \label{eq: BtoA}
 t_2 + x(t_2) = L
 \end{equation}
Let the corresponding instantaneous rest frame be ${\rm S}_2$. Simultaneously with ${\rm A}_2$ relative to ${\rm S}_2$, the emitting astronaut is at the event ${\rm B}_2 = \left( t_2 + L \gamma(v_2) v_2,x(t_2) + L \gamma(v_2),0,0 \right)$. The signal arrives to A blue-shifted with the Doppler factor
 \begin{equation} \label{eq: DoppBA}
 \sqrt{\frac{1+v_2}{1-v_2}} = e^{\eta(v_2)} = \exp{\left[\frac{\tau\left({\rm B}_o{\rm B}_2\right) - \tau\left({\rm A}_o{\rm A}_2\right)}{L}\right]}
 \end{equation}
Here, again, the proper-times lapses are respectively computed on the world-line intervals ${\rm B}_o{\rm B}_2$ and ${\rm A}_o{\rm A}_2$.

\begin{figure}[h]
  \centering
  \includegraphics[width=6cm]{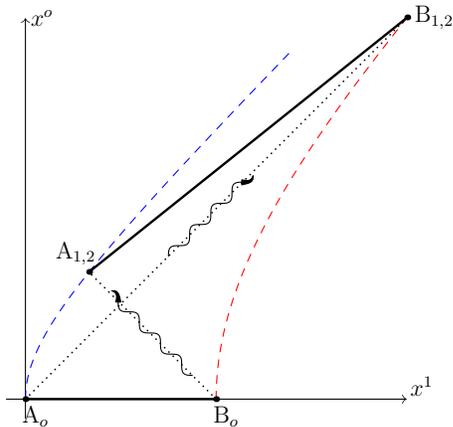}
  \begin{minipage}[b]{250pt}
  \caption{\label{fig: lighthyp} Space-time diagram showing light signals sent from A (${\rm A}_o$) to B (${\rm B}_{1,2}$) and from B (${\rm B}_o$) to A (${\rm A}_{1,2}$). In this diagram, which depicts hyperbolic motion, the signals emitted simultaneously also arrive simultaneously relative to the spaceship's instantaneous rest-frame.}
  \end{minipage}
\end{figure}

\vskip20pt

\section{Simultaneity of signal transmission}

The symmetry between the above relations, in particular equations \eref{eq: DoppAB} and \eref{eq: DoppBA}, is evident, raising the question whether, if the signals are emitted simultaneously (as at the events ${\rm A_o}$ and ${\rm B_o}$) relative to the momentary rest frame, will they also arrive simultaneously relative to an instantaneous rest frame ? In other words, in the above notation, is it possible that there is some $t$ so that $t_1 = t_2 = t$, $v_1 = v_2= v(t)$, and both frames ${\rm S}_1$ and ${\rm S}_2$ coincide ?

The answer depends on the details of the journey, namely on the function $x(t)$ together with the derived $v(t)$. Since the spaceship accelerates it is convenient to use, as long as $a \ne 0$ (which is assumed in the following), the rapidity $\eta\left( v \right) \equiv  \tanh^{-1} \left( v \right)$ as time-like evolution parameter. Using the relations $\gamma \left( v \right) = \cosh \eta$ and $d\tau = \gamma^{-1}\left(v\right) dt = dt/\cosh \eta$ for A's proper-time together with $d\tau = d\eta / a$ (eq. \eref{eq: tauaeta}) then follow the relations
\begin{equation}\label{eq: dtdx}
 \eqalign{
 dt &= \cosh \eta d\tau = \frac{\cosh \eta d\eta}{a} \, , \\
 dx &= \tanh \eta dt = \sinh \eta d\tau = \frac{\sinh \eta d\eta}{a} }
\end{equation}
Assuming (without loss of generality) the initial conditions $t = 0, \, v(0) = v_0$ with $\eta_0 = \eta\left(v_0\right)$ and $x(0) = 0$ for the signals' simultaneous emission (taking into account also the possibility that $v_0, \eta_0 \ne 0$, so that the spaceship is not necessarily at rest relative to S at the moment of emission), the integral relations ensue,
\begin{equation}\label{eq: inttx}
 t\left( \eta \right) = \int\limits_{\eta_0}^\eta {\frac{\cosh \eta}{a\left( \eta \right)} d\eta} \quad  ,  \quad x\left( \eta \right) = \int\limits_{\eta_0}^\eta {\frac{\sinh \eta}{a\left( \eta \right)} d\eta}
\end{equation}

The light-like conditions for the simultaneous arrival of the signals on some hyperplane corresponding to $t = t_1$ (connecting $x_{\rm A}^\mu(t_1)$ with $x_{\rm B}^\mu(t_1)$) with rapidity $\eta_1 = \eta\left(v(t_1)\right)$ were found (eqs. (22) \& (24), now with $\eta_0$ possibly non-zero) to be
 \begin{equation} \label{eq: llconds}
 \eqalign{
 &{\rm A} \to {\rm B} : \quad t_1 - x\left(t_1\right) = e^{- \eta_1} L \\
 &{\rm B} \to {\rm A} : \quad t_1 + x\left(t_1\right) = e^{\eta_0} L }
 \end{equation}
Therefore, using \eref{eq: inttx}, the condition for simultaneous arrival of the signals on the hyperplane corresponding to $\eta = \eta_1$ is the existence of a common solution $\left(\eta_0,\eta_1\right)$ to
\begin{equation}\label{eq: simcond1}
 \int\limits_{\eta_0}^{\eta_1} {\frac{e^{\eta_1 - \eta}d\eta}{a\left( \eta \right)}} = \int\limits_{\eta_0}^{\eta_1} {\frac{e^{\eta - \eta_0} d\eta}{a\left(\eta \right)}} = L \, .
\end{equation}

Once the proper acceleration $a\left(\eta\right)$ is known, both $\eta_0$ and $\eta_1$ are determined by the set of double equations in \eref{eq: simcond1}. Defining $\bar\eta = \left(\eta_0 + \eta_1\right)/2$ and $\Delta\eta = \left(\eta_1 - \eta_0\right)/2$, the left equality in \eref{eq: simcond1} implies that a necessary and sufficient condition for simultaneous arrival of the signals is the existence of a non-trivial solution, with $\Delta \eta \ne 0$, of the equation
\begin{equation}\label{eq: simcondf}
 \int\limits_{\bar\eta - \Delta\eta}^{\bar \eta + \Delta\eta} {\frac{{\sinh \left( \eta - \bar\eta \right)d\eta}}{a\left( \eta \right)}} = 0
\end{equation}
for some $\bar\eta$ (clearly, $\Delta\eta = 0$ necessarily implies $L = 0$).

The rapidity has the property that under a Lorentz transformation in 1+1 dimensions it changes by an additive constant. The difference $\Delta\eta$ is therefore a Lorentz-invariant quantity. Let $\bar{\rm{S}}$ be an inertial frame moving relative to S with velocity $V = \tanh\bar\eta$. If $\eta$ is the momentary rapidity of the spaceship relative to S then $\eta' = \eta - \bar\eta$ is the rapidity of the spaceship, at the same moment, relative to $\bar{\rm{S}}$. In more familiar terms, if $v = \tanh\eta$ is the momentary velocity of the spaceship relative to S then $v' = \tanh\eta' = \tanh\left(\eta - \bar\eta\right)$ is the velocity of the spaceship at the same moment relative to $\bar{\rm{S}}$. In particular, $\bar{\rm{S}}$ is the momentary rest-frame of the spaceship when the latter moves with velocity $V$ relative to S.

In terms of $\eta'$ eq.\eref{eq: simcondf} becomes
\begin{equation}\label{eq: simcondf2}
 \int\limits_{-\Delta\eta}^{\Delta\eta} {\frac{\sinh \left(\eta'\right)d\eta'}{a\left(\eta' + \bar\eta\right)}} = 0
\end{equation}
The proper acceleration is Lorentz invariant, but as a function of the rapidity, which is frame-dependent, we must be careful with the reference frame it is associated with. The function $a\left(\eta\right) = a\left(\eta' + \bar\eta\right)$, which appears in all the equations so-far is associated with reference frame S because its argument, the rapidity $\eta$, is measured relative to S, and should appropriately be denoted $a_{\rm{S}}\left(\eta\right)$. We may alternatively define $a_{\bar{\rm{S}}}\left(\eta'\right)=a_{\rm{S}}\left(\eta' + \bar\eta\right)$ as the proper acceleration function associated with $\bar{\rm{S}}$, and write \eref{eq: simcondf2} as
\begin{equation}\label{eq: simcondf3}
 \int\limits_{-\Delta\eta}^{\Delta\eta} {\frac{\sinh \left(\eta'\right)d\eta'}{a_{\bar{\rm{S}}}\left(\eta'\right)}} = 0
\end{equation}
Obviously the two equations are equivalent, but \eref{eq: simcondf3} has a more symmetric appearance.

In particular, if the acceleration has the symmetry property that
\begin{equation}\label{eq: asymcond}
 a_{\bar{\rm{S}}}\left(\eta'\right) = a_{\bar{\rm{S}}}\left(-\eta'\right) \quad \Leftrightarrow \quad a_{\rm{S}}\left(\eta\right) = a_{\rm{S}}\left(2\bar\eta - \eta\right)
\end{equation}
then the simultaneity condition \eref{eq: simcondf3} is automatically satisfied for all $\Delta\eta$. $\Delta\eta$ is then determined, dependent on $L$, by \eref{eq: simcond1}.

In the case of hyperbolic motion with constant acceleration $a$, \eref{eq: simcondf3} is automatically satisfied for {\it all} $a$, $\bar\eta$ and $\Delta\eta$, and integration of \eref{eq: simcond1} yields $e^{2\Delta\eta} = 1 + aL$. A counter example is suggested by $x(t) = \alpha t^2$, $\alpha$ being some arbitrary constant. The proper acceleration here is $a_{\rm{S}}(\eta) = 2 \alpha \cosh^3\eta$, and substitution in \eref{eq: simcondf} yields
\begin{equation}\label{eq: }
 \int\limits_{\bar\eta - \Delta\eta}^{\bar\eta + \Delta\eta} {\frac{\sinh\left(\eta - \bar\eta\right)d\eta}{\cosh^3\eta}} = -\frac{\sinh\bar\eta \sinh^2\left( \Delta\eta \right)\sinh\left( 2\Delta\eta \right)}{\cosh^2\left( \bar\eta + \Delta\eta \right) \cosh^2\left( \bar\eta - \Delta\eta \right)} = 0 \, ,
\end{equation}
so the simultaneity condition is satisfied only for $\bar\eta = 0$ or $\Delta\eta = 0$. If it is assumed in this particular example that the motion relative to the S-frame is only in the $+x$-direction, so that $\eta \ge 0$, then a solution exists only for the trivial solution $\Delta\eta = 0$. This counter-example demonstrates that for a non-trivial solution it must be verified that the solution of eqs. \eref{eq: simcond1} exists within the allowed $\eta$-domain.

The physical meaning of these results is as follows : The spaceship arrives to state of rest relative to $\bar{\rm{S}}$ after the emission of the signals, but before their arrival to their targets. From the standpoint of $\bar{\rm{S}}$-observers, when the signals are emitted the spaceship is moving to the left with velocity $v_1 = -\tanh\Delta\eta$, then it slows down, comes to rest and then starts moving to the right, accelerating. The signals' arrival is when the spaceship's velocity relative to $\bar{\rm{S}}$ is $v_2 = \tanh\Delta\eta$. Because of the relative motion, neither the emission nor the arrival of the signals are simultaneous for the $\bar{\rm{S}}$-observers: Comparison of \eref{eq: xA} and \eref{eq: xB} relative to $\bar{\rm{S}}$, or Lorentz transforming from the spaceship's momentary rest frames to $\bar{\rm{S}}$, verifies that in terms of $\bar{\rm{S}}$-time $\bar t$, the emission from A precedes the emission from B with $\Delta\bar t_{1} = \bar t_{1\rm{A}} - \bar t_{1\rm{B}} = L\sinh\Delta\eta$, while the arrival to B precedes the arrival to A by the same amount, $\Delta\bar t_{2} = \bar t_{2\rm{B}} - \bar t_{2\rm{A}} = L\sinh\Delta\eta$. In other words, for the $\bar{\rm{S}}$-observers, the signal from B to A seems to be emitted and arriving before the corresponding events for the signal from A to B.

\vskip20pt

\section{Concluding remarks -- on age and proper-time measurement in relativistic extended systems} \label{sec: conrem}

This article offers a platform for a combined discussion of two of the fundamental issues of special relativity -- the association of proper-time with age and relativistic rigid motion, while using the concepts inherent in the so called `twin paradox'.  Apart from the anecdotical aspect of suggesting and discussing a non-standard version for the `twin paradox', this note emerged from a work which addressed the question of {\it Whether it is possible to assign the concept of common proper-time to complex, spatially extended, relativistic systems as a whole}; in particular, with the wish to use this common proper-time for the {\it age} of the system.

For a pointlike body, the proper-time measurement is identical with the reading of a clock momentarily at rest with the body : An un-accelerated point particle may always be found at rest relative to some inertial frame, so the proper-time measurement for it is identical to the clock reading in that frame. Otherwise, if accelerated, the proper-time lapse of the particle moving on the world-line $\left(t,\vec r (t) \right)$ is the integral $\int\sqrt{dt^2 - d\vec r\,^2} = \int\sqrt{1-\vec v\,^2} dt$ along its world-line. Since this is the only time measurement available for that particle, it must necessarily serve as the measure for its {\it age}.

Regarding the space mission, the twin astronauts may be regarded as pointlike bodies, while the spaceship was deliberately considered rigid to insure that there is always an inertial frame in which the whole spaceship is momentarily at rest. Still, since Lorentz transformation of time depends on the location, different parts of the system measure different proper-time lapses.

Extended systems consist of number of points, each defining a different world-line. Even if the whole system may be found momentarily, simultaneously, at rest, still different proper-times, different ages, are measured at different points. Also, it is not possible to identify, at least not from kinematical considerations alone, a preferred point which may serve to define the common proper-time for the whole system : \Eref{eq: aa-ab2} verifies that in this sense all the points of the system are equivalent.

It is noted that the present version of the `twin paradox', which really is not a paradox at all, is closely related to another famous relativity `paradox', the so-called `Bell's spaceships paradox' \cite{BellSSP} (which is also not a paradox at all). To fit our story with Bell's we could have used, instead of the long spaceship, two small spaceships connected by a long rigid rod, but this is an un-essential difference. Bell's `paradox' was recently discussed by Franklin \cite{Franklin2010}, who, among other things, also compared the Minkowskian times of the right and left spaceships (or brothers) which are obviously the same in any instantaneous rest frame. However, the {\it ages} of the brothers are determined not by the Minkowskian times but by the proper-times measured along their (separate) space-time trajectories. As we have seen, the result is that if the end station is moving relative to the home station, then the brothers do indeed end up with different ages, simply because of being located in remote parts of the spaceship.

This marks a difference between the existence of momentary rest frames for the whole (extended) system, on the one hand, and the possibility of assigning a common proper time for the whole system, on the other hand.

We end with a comment regarding the significance of assuming rigid motion in this type of `twins paradox' scenario. The rigidity condition implies, by definition, that the relative distance between the space travelers be maintained constant relative to themselves. The requirement to maintain constant relative distance is introduced to provide means to compare the ages -- proper-times -- of the twins after the journey. These distances are compared between events on the twins' world-lines that are simultaneous in some inertial reference frame. In the rigidity condition these are the twins' momentary rest frames, but in principle it is possible to envisage using also other frames for the simultaneity.

As an example, a similar but different `twin paradox' scenario was proposed by Boughn \cite{Boughn1989}. In this scenario the two twins sit in point-like spaceships, initially at rest in some inertial reference frame S, separated distance $L$ apart. They start together their journey and supposed to follow the same plan relative to the home-base, so their world-lines, in S coordinates, are
 \begin{equation} \label{eq: xAxB.B}
 x_{\rm A}^\mu = \left(t,x(t),0,0 \right) \qquad , \qquad  x_{\rm B}^\mu = \left(t,x(t) + L,0,0 \right)
\end{equation}
where $x(t)$ describes some non-uniform motion.

Here the requirement that the twins maintain constant relative distance is assumed to hold simultaneously relative to the home-base reference frame S, which is a fixed inertial frame. But for the astronauts themselves this simultaneity is irrelevant -- they can only measure distances, velocities and accelerations relative to themselves, {\it i.e.}, in an inertial frame in which they are momentarily at rest. Therefore, for twin A at the event $x_{\rm A}^\mu (t) = \left(t,x(t),0,0 \right)$, the simultaneous B-event in the scenario \eref{eq: xAxB.B} is not $x_{\rm B}^\mu (t)$ but rather $x_{\rm B}^\mu \left(\bar t\right) = \left(\bar t,x\left(\bar t\right) + L,0,0 \right)$ determined by the simultaneity condition
\begin{equation}\label{simcondB}
 \fl \hskip20pt \left[ {x_{\rm B}^\mu \left( \bar t \right) - x_{\rm A}^\mu \left( t \right)} \right] \cdot \frac{dx_{\rm A}^\mu \left( t \right)}{dt} = 0  \quad \Rightarrow \quad  \bar t - t = \left[ x\left( \bar t \right) + L - x\left( t \right) \right] \cdot v\left( t \right) \, .
\end{equation}
Since $\bar t \ne t$ then also $v\left(\bar t\right) \ne v(t)$, each twin sees the other moving, and actually being accelerated relative to him/her. The only way that the relative distance may be regarded constant by the twins relative to themselves is, therefore, in rigid motion.

\vskip20pt

\rule{10cm}{1pt}


  \end{document}